\documentclass[12pt]{article}
\usepackage{amsmath,amsfonts,amssymb,epsfig,subfigure}
\usepackage{amsmath,amsfonts,amssymb}

\usepackage{color}

\renewcommand{\le}{\leqslant}
\renewcommand{\ge}{\geqslant}
\newcommand{\be}{\begin{equation}}
\newcommand{\en}{\end{equation}}

\newcommand{\imi}{\textrm{i}}
\renewcommand{\vec}[1]{\boldsymbol{#1}}

\begin{document}
\numberwithin{equation}{section}

\title{Bleustein-Gulyaev waves in some \\ functionally graded
materials}
\author{
  Bernard Collet, 
  Michel Destrade, 
  G\'erard A.~Maugin}
\date{2006}
\maketitle


\begin{abstract}

Functionally Graded Materials are inhomogeneous elastic bodies whose
properties vary continuously with space. 
Hence consider a half-space ($x_2>0$) occupied by a special 
Functionally Graded Material made of an hexagonal (6mm)
piezoelectric crystal for which the elastic stiffness $c_{44}$, 
the piezoelectric constant $e_{15}$, 
the dielectric constant $\epsilon_{11}$, and the
mass density, all vary proportionally to the same ``inhomogeneity
function'' $f(x_2)$, say. 
Then consider the problem of a piezoacoustic
shear-horizontal surface wave which leaves the interface ($x_2 = 0$) 
free of mechanical tractions and vanishes as $x_2$ goes to infinity 
(the Bleustein-Gulyaev wave). 
It turns out that for some choices of the
function $f$, this problem can be solved exactly for the usual boundary
conditions, such as metalized surface or free surface. 
Several such functions $f(x_2)$ are derived here, such as
$\exp(\pm 2\beta x_2)$ ($\beta$ is a constant) which is 
often encountered in geophysics, 
or other functions which are periodic or which vanish as 
$x_2$ tends to infinity; 
one final example presents the advantage of describing
a layered half-space which becomes asymptotically homogeneous away from
the interface. 
Special attention is given to the influence of the
different inhomogeneity functions upon the characteristics of the
Bleustein-Gulyaev wave (speed, dispersion, attenuation factors, depth
profiles, electromechanical coupling factor, etc.) 

\end{abstract}

\section{Introduction}

The wireless communication industry (mobile phones, global positioning 
systems, pagers, label identification tags, etc.) fuels most of the 
current mass production of Surface Acoustic Wave devices (more than 1 
billion units/year) where SAW-based interdigital transducers are used 
as  high-frequency filters.
In the race for miniaturization, devices based on Bleustein-Gulyaev 
waves (pure shear-horizontal mode) technology have proved more apt 
for downsizing than those based on Rayleigh waves 
(two- or three-partial modes) technology, according to Kadota et al.
(2001).

Can the so-called ``Functionally Graded Materials'', whose properties
vary continuously in space, be used to improve 
the efficiency of Bleustein-Gulyaev waves? 
For a 6mm piezoelectric \textit{homogeneous substrate}, the classic 
solution of Bleustein (1968) and Gulyaev (1969) 
is quite simple to derive; 
for a \textit{functionally graded substrate}, the corresponding wave 
solution is in general impossible to determine analytically. 
In order to make progress, and with a view to use the eventual results 
as benchmarks for more complicated simulations, 
this paper strikes a compromise between these two extreme situations, 
and aims at finding in a simple way certain types of functionally 
graded substrates for which analytical Bleustein-Gulyaev type of 
solutions are easily derived. 
Such a task can be achieved by making the assumption that for the 
functionally graded material, the elastic stiffness $c_{44}$, 
the piezoelectric constant $e_{15}$, the dielectric constant 
$\epsilon_{11}$, \textit{and} the mass density $\rho$, all vary in 
the same proportion with a single space variable.
This assumption is often encountered in the literature, 
see for example the recent articles 
(Jin et al., 2003; Kwon and Lee, 2003; Wang, 2003; 
Chen et al., 2004; Kwon, 2004; Ma et al., 2004; 
Chen and Liu, 2005$a,b$; Guo et al., 2005$a,b$; Ma et al., 2005$a,b$;
Pan and Han, 2005; Sladek et al., 2005; Sun et al., 2005; 
Feng and Su, 2006). 
It is a strong assumption, which can be envisaged to hold for 
$c_{44}$,  $e_{15}$,  $\epsilon_{11}$ in certain contexts 
(pre-stressed laminae (Cohen and Wang, 1992), 
elastic bodies subjected to a thermal gradient (Saccomandi, 1999),
continuously twisted structurally chiral media 
(Lakhtakia, 1994), etc.), but is unlikely to hold for $\rho$ as well. 
However, this assumption proves crucial for the derivation of 
analytical results in terms of ``simple'' functions such as the 
polynomial, sinusoidal, and hyperbolic functions. 
Previous studies have indeed shown  
that if $\rho$ behaves differently from the 
other material quantities, then analytical solutions of the 
shear-horizontal wave problem involve special functions such as Bessel 
functions (Wilson, 1942; Bhattacharya, 1970; Maugin, 1983), 
Hankel functions (Deresiewicz, 1962), 
Whittaker functions (Deresiewicz, 1962; Bhattacharya, 1970), 
hypergeometric functions (Bhattacharya, 1970; Viktorov, 1979;
Maugin, 1983), etc. 
(see the review by Maugin (1983) for some pointers to the 
wide literature on the subject.);
otherwise, numerical and approximate methods are necessary to 
solve the problem, such as those based on Laguerre series 
(Gubernatis and Maradudin, 1987), 
on a combination of Fast Fourier Transforms and modal 
analysis (Liu and Tani, 1994), 
on Legendre polynomials (Lefebvre et al., 2001), 
on the WKB approximation (Liu and Wang, 2005), etc.

In short, some generality is lost by taking $\rho$ to behave in the 
same manner as the other quantities, but some simplicity and 
insights are gained, because the resulting exact solutions may 
serve as benchmarks for more realistic situations, 
where for instance 
a perturbation boundary element method can be used 
(Azis and Clements, 2001). 
The governing equations derived in the course of this paper can also 
be specialized to the consideration of anti-plane deformations in the 
context of piezo-elastostatic problems, 
where the density plays no role and its 
eventual spatial variations need not be specified.

The paper begins the analysis in Section 2 with the derivation of 
the equations governing the propagation of a Shear-Horizontal 
wave in the type of Functionally Graded Material just discussed. 
In Section 3, a change of unknown functions leads to the decoupling of 
the four first-order governing equations into  two separate pairs of  
second-order differential equations. 
For certain choices of inhomogeneity, the differential equations
have constant coefficients and the consequences of such choices 
on the propagation of Bleustein-Gulyaev waves are fully analyzed and 
are illustrated numerically by two examples:
one where the inhomogeneity is a decreasing exponential function, 
the other where it is an inverse quadratic function. 
The last two sections show that other inhomogeneity functions leading 
to explicit results can be generated, not necessarily by seeking 
differential equations with constant coefficients. 
Section 4 focuses on an inhomogeneity function for which the material 
parameters vary smoothly from a value at the interface to an 
asymptotic value at infinite distance from the interface. 
Section 5 presents one method, presumably among many others, to 
generate an infinity of inhomogeneity functions leading to 
exact Bleustein-Gulyaev solutions.

\section{A certain type of functionally graded materials}

The Bleustein-Gulyaev wave is a shear horizontal wave, travelling over 
the surface of a semi-infinite piezoelectric solid for which the 
sagittal plane is normal to a binary axis of symmetry.
Now consider a half-space $x_2 \ge 0$ (say), made of a piezoelectric 
crystal with 6mm symmetry (see e.g. Royer and Dieulesaint, 2000) 
and with continuously varying properties in the $x_2$-direction.
Specifically, the elastic stiffness $c_{44}$, the piezoelectric 
constant $e_{15}$, the dielectric constant $\epsilon_{11}$, and the 
mass density $\rho$, all vary in the same proportion with depth $x_2$:
\begin{equation} \label{inhomogeneityLAw}
 \{ c_{44}(x_2), e_{15}(x_2), \epsilon_{11}(x_2), \rho(x_2) \}  
  =  \{ c^\circ_{44}, e^\circ_{15}, \epsilon^\circ_{11}, \rho^\circ \}
          f(x_2),
\end{equation}
where $ c^\circ_{44}$, $e^\circ_{15}$, $\epsilon^\circ_{11}$, 
$\rho^\circ$ are constants, and $f$ is a yet unspecified 
function of $x_2$, henceforward called the 
\textit{inhomogeneity function}.
Without loss of generality, $f$ is normalized as $f(0)=1$.

Now take two orthogonal directions $x_1$, $x_3$ in the 
plane $x_2=0$ such that the symmetry axis is along $x_3$ and 
consider the propagation of a Bleustein-Gulyaev wave,
traveling with speed $v$ and wave number $k$ in the $x_1$-direction.
The associated quantities of interest are: the mechanical displacement 
component $u_3$, the electric potential $\phi$, the mechanical 
traction components $\sigma_{13}$, $\sigma_{23}$, and the electric 
displacement components $D_1$, $D_2$. 
They are taken in the form
\begin{equation} \label{wave}
\{ u_3, \phi, \sigma_{j3}, D_j \}(x_1, x_2, t)
  = \{ U_3(x_2), \varphi(x_2), \imi t_{j3}(x_2), \imi d_j(x_2)\}
                \text{e}^{\imi k(x_1 - vt)},
\end{equation}
where $U_3$, $\varphi$, $t_{j3}$, $d_j$ ($j = 1,2$) are unknown 
functions of $x_2$ alone, to be determined from the piezoacoustic 
equations and from the boundary conditions.

In the present context, the classical equations of piezoacoustics
written in the quasi-electrostatic approximation, 
\begin{equation}
\partial \sigma_{ij} / \partial x_j = 
  \rho \; \partial^2 u_i / \partial t^2, \quad 
\partial D_j / \partial x_j = 0,
\end{equation}
decouple entirely the anti-plane stress and strain from their in-plane 
counterparts. 
The anti-plane equations can be written as a first-order differential 
system,
\begin{equation} \label{1stOrder}
   \begin{bmatrix} \vec{u}' \\ \vec{v}' \end{bmatrix}
     = \imi \begin{bmatrix}
                    \vec{0} & \dfrac{1}{f(x_2)}\vec{N}_2 \\
                     k^2f(x_2) \vec{K}  & \vec{0}
                             \end{bmatrix}
   \begin{bmatrix} \vec{u} \\ \vec{v} \end{bmatrix},
    \quad \text{where} \quad
  \vec{u} :=   \begin{bmatrix} U_3 \\ \varphi \end{bmatrix},
  \vec{v} :=   \begin{bmatrix} t_{23} \\ d_2 \end{bmatrix},
 \end{equation}
and $\vec{N}_2$, $\vec{K}$ are the following constant symmetric 
matrices,
\begin{equation}  \label{N2Kdef}
 \vec{N}_2 := 
    \dfrac{1}{c^\circ_{44} \epsilon^\circ_{11} + e^{\circ 2}_{15}}
 \begin{bmatrix}
       \epsilon^\circ_{11} & e^\circ_{15} \\
       e^\circ_{15}& -c^\circ_{44}
       \end{bmatrix},
\quad
 \vec{K} := \begin{bmatrix}
       \rho^\circ v^2 - c^\circ_{44} & -e^\circ_{15}\\
           -e^\circ_{15} & \epsilon^\circ_{11} 
      \end{bmatrix}.
\end{equation}

\section{Some simple inhomogeneity functions}

In this Section, attention is restricted to some inhomogeneity 
functions for which the piezoacoustic equations turn into linear 
ordinary differential equations with \textit{constant} 
coefficients.

\subsection{Further decoupling of the piezoacoustic equations}

With the new vector functions $\vec{\hat{u}}$ and $\vec{\hat{v}}$, 
defined as
\begin{equation} \label{scaling}
\vec{\hat{u}} (x_2) = \sqrt{f(x_2)} \vec{u}(x_2), \quad 
\vec{\hat{v}} (x_2) = \vec{v}(x_2)/\sqrt{f(x_2)}, 
\end{equation}
the system  \eqref{1stOrder} becomes
\begin{equation} \label{scaledSystem}
   \begin{bmatrix} \vec{\hat{u}}' \\ \vec{\hat{v}}' \end{bmatrix}
     = \begin{bmatrix}
                    \frac{p}{2}\vec{1} & \imi \vec{N}_2 \\
                     \imi k^2 \vec{K}  & -\frac{p}{2}\vec{1}
                             \end{bmatrix}
   \begin{bmatrix} \vec{\hat{u}} \\ \vec{\hat{v}} \end{bmatrix},
    \quad \text{where} \quad
  p :=   \dfrac{f'}{f}.
 \end{equation}
Now, by differentiation and substitution, an entirely decoupled 
second-order system emerges:
\begin{equation} \label{2ndSystem}
   \begin{bmatrix} \vec{\hat{u}}'' \\ \vec{\hat{v}}'' \end{bmatrix}
     = -\begin{bmatrix}
   k^2 \vec{N_2K} - (\frac{p^2}{4} + \frac{p'}{2})\vec{1} & \vec{0} \\
   \vec{0}  &  k^2 \vec{KN_2} - (\frac{p^2}{4} - \frac{p'}{2})\vec{1}
       \end{bmatrix}
   \begin{bmatrix} \vec{\hat{u}} \\ \vec{\hat{v}} \end{bmatrix},
 \end{equation}
and two simple ways of finding exact solutions for shear-horizontal 
wave propagation appear naturally. 
 
$\bullet$ Either \textit{(i)} solve
 \begin{equation} \label{(i)}
  \dfrac{p^2}{4} + \dfrac{p'}{2} = c_0,
 \end{equation}
where $c_0$ is a constant. Then the solution $\vec{\hat{u}}$ to the 
second-order equation \eqref{2ndSystem}$_1$ with (now) 
\textit{constant} coefficients is easily found.
Finally, $\vec{u}$ follows from \eqref{scaling}$_1$ 
and $\vec{v}$ from the inversion of \eqref{1stOrder}$_1$, 

$\bullet$ Or \textit{(ii)} solve
 \begin{equation} \label{(ii)}
  \dfrac{p^2}{4} - \dfrac{p'}{2} = c_0,
 \end{equation}
where $c_0$ is a constant. 
Then the solution $\vec{\hat{v}}$ to the 
second-order equation \eqref{2ndSystem}$_2$ with (now) 
\textit{constant} coefficients is easily found. 
Finally, $\vec{v}$ follows from \eqref{scaling}$_2$ 
and $\vec{u}$ from the inversion of \eqref{1stOrder}$_2$. 

Of course the two possibilities \eqref{(i)} and \eqref{(ii)} do not 
exhaust the classes of solutions. 
The last section of this article shows how infinitely more 
inhomogeneous profiles can be generated.

Clearly now, if $p$ is solution to \eqref{(i)}, then $-p$ is solution 
to \eqref{(ii)}; so that if  $f$ is solution to \eqref{(i)}, then 
$1/f$ is solution to \eqref{(ii)}.
The resolution of these two equations is straightforward, 
and the results are collected in Table 1: 
according as to whether $c_0$ is positive, negative, 
or equal to zero (second column), 
several functions $p(x_2)$ (third column) 
and $f(x_2)$ (fourth column) are found. 
The inhomogeneity 
profile P0 is common to the resolution of \eqref{(i)} and 
\eqref{(ii)}; 
profiles P1-P5 result from \eqref{(i)} and P6-P10 from \eqref{(ii)}. 
The quantities $\beta$ (inverse of a 
length) and $\delta$ (non-dimensional) are arbitrary, so that the 
inhomogeneity functions $f(x_2)$ in P4 and P5 are essentially the same 
functions, and so are the inhomogeneity functions in P9 and P10.

Some of these inhomogeneity functions are often encountered in the 
geophysics literature, such as the exponential function P0 or the 
quadratic function P1. 
Dutta (1963) used P2 for Love waves; Erdogan and Ozturk 
(1992) and Hasanyan et al. (2003) derived P0-P5 in a different 
(purely elastic) context; 
P6-P10 appear to be new, presumably because the preferred 
second-order form of the equations of motion is usually 
\eqref{2ndSystem}$_1$ rather than \eqref{2ndSystem}$_2$ 
(see (Destrade, 2001) for a discussion on this latter point.)

The  functions found present the advantages of 
mathematical simplicity and familiarity. 
Each of them however presents the inconvenience of describing a 
somewhat unrealistic inhomogeneity, because each either blows up or 
vanishes as $x_2 \rightarrow \infty$, or blows up or vanishes 
periodically. 
These problems can be overcome by considering that they occur 
sufficiently far away from the interface, and by focusing on the 
near-the-surface localization of the wave.

\subsection{Exact solution}

Here the emphasis is on the complete resolution for the 
Bleustein-Gulyaev wave in Case \textit{(i)} (profiles P0-P5). 
In Case \textit{(ii)}, the resolution is very similar, and the 
corresponding results are summarized at the end of this subsection.

First, solve the decoupled, second-order, linear, with 
constant coefficients, differential equation 
 \eqref{2ndSystem}$_1$  for $\vec{\hat{u}}$:
\begin{equation} \label{decoupled}
 \vec{\hat{u}}'' + 
    (k^2 \vec{N_2K} - c_0\vec{1}) \vec{\hat{u}} = \vec{0},
\end{equation}
with a solution in exponential evanescent form,
\begin{equation}
 \vec{\hat{u}}(x_2) = \text{e}^{-k q x_2} \vec{\hat{U}^0}, 
 \quad 
 \Re(q) >0,
\end{equation}
where $\vec{\hat{U}^0}$ is constant and $q$ is an attenuation factor. 
Then $\vec{\hat{U}^0}$  and $q$ are solutions to 
\begin{equation} \label{eigen}
    [k^2 \vec{N_2K} - (c_0 - k^2 q^2) \vec{1}] \vec{\hat{U}^0} 
         = \vec{0}.
\end{equation}
The associated determinantal equation is the 
\textit{propagation condition}, here
\begin{equation}
 [ k^2(q^2-1 - (v/v_T^\circ)^2 - c_0][ k^2(q^2-1) - c_0]  = 0,
\end{equation}
where $v_T^\circ$ is the speed of the bulk shear wave in the 
homogeneous ($f \equiv 1$) material, given by 
\begin{equation}
\rho^\circ  v_T^{\circ 2} = 
   c_{44}^\circ + e_{15}^{\circ 2} / \epsilon_{11}^\circ.
\end{equation} 

The \textit{attenuation factors} $q_1$, $q_2$ (say) with positive 
real part are 
\begin{equation} \label{q1q2}
q_1 = \sqrt{1 + c_0 / k^2 - (v/v_T^\circ)^2}, \quad 
 q_2 = \sqrt{1 + c_0 / k^2},
\end{equation} 
provided the speed belongs to the subsonic interval 
\begin{equation} \label{condition}
0 < (v/v_T^\circ)^2 < 1 + c_0 / k^2.
\end{equation}
The smallest of these two quantities ($q_1$) is indicative of the 
\textit{penetration depth}. 
Here, the inhomogeneity affects the penetration depth in the following 
manners: for the exponential profile P0 and for the 
hyperbolic profiles (P2, P3, P7, P8), the wave 
is more localized than in the homogeneous case ($f \equiv 1$); 
for the trigonometric profiles (P4, P5, P9, P10), the wave 
penetrates further into the substrate; 
for the polynomial profiles (P1, P6), the penetration depth is the 
\textit{same} as in the homogeneous case.
Note in passing that the inequality \eqref{condition} puts an 
upper bound on the possible values of $\beta$ for the  
trigonometric profiles P4, P5, P9, P10 
(where $c_0 = -\beta^2$), namely:
$\beta^2 < k^2$, which means that the wavelength of the wave 
must be smaller than the wavelength of those profiles.
These remarks are however preliminary and concern the behavior of the 
functions $\vec{\hat{u}}$ and $\vec{\hat{v}}$ with depth. 
The behavior of the wave itself is dictated by the functions 
$\vec{u}$ and $\vec{v}$, see \eqref{scaling}.
In particular, inequality \eqref{condition} ensures that 
$\vec{\hat{u}}(\infty) = \vec{0}$ but, because 
$\vec{u} = (1/\sqrt{f})\vec{\hat{u}}$, not necessarily that 
$\vec{u}(\infty) = \vec{0}$; this latter condition must be tested 
a posteriori against each different form of $f$.

Now the constant vectors $\vec{\hat{U}^1}$, $\vec{\hat{U}^2}$ (say) 
satisfying \eqref{eigen} when $q = q_1$, $q_2$, respectively, 
are easily computed and the general solution to \eqref{decoupled}
is constructed as: 
$\vec{\hat{u}}(x_2) = \gamma_1  \text{e}^{-k q_1 x_2} \vec{\hat{U}^1}
  + \gamma_2  \text{e}^{-k q_2 x_2} \vec{\hat{U}^2}$ where $\gamma_1$, 
  $\gamma_2$ are constant scalars. 
Explicitly,
\begin{equation}
\vec{\hat{u}}(x_2) = 
  \gamma_1  \text{e}^{-k q_1 x_2} 
     \begin{bmatrix} 
        1 \\ \frac{e_{15}^\circ}{\epsilon_{11}^\circ}
     \end{bmatrix}
  + \gamma_2  \text{e}^{-k q_2 x_2} 
        \begin{bmatrix} 
          0 \\ 1
        \end{bmatrix}.
\end{equation}
Then $\vec{u}$ follows from \eqref{scaling}$_1$ as: 
$\vec{u} = (1/\sqrt{f}) \vec{\hat{u}}$, and $\vec{v}$ 
follows from the substitution of this latter equation into 
the inverse of \eqref{1stOrder}$_1$, which is: 
$\vec{v} = - \imi f  \vec{N_2}^{-1} \vec{u}'$. 
In the end, it is found that at the interface,
\begin{align}
& U_3(0) =  \gamma_1, 
\notag \\
& \varphi(0) =  \dfrac{e_{15}^\circ}{\epsilon_{11}^\circ}\gamma_1 
                   + \gamma_2,
\notag \\
& t_{23}(0) = \imi k
    \left[\left( c^\circ_{44} 
          + \dfrac{e_{15}^{\circ 2}}{\epsilon_{11}^\circ}\right)
     \left(q_1 + \dfrac{f'(0)}{2k} \right)\gamma_1 
    + e_{15}^\circ\left(q_2 + \dfrac{f'(0)}{2k} \right)\gamma_2
      \right] ,
\notag \\
& d_2(0) = -\imi k \epsilon_{11}^\circ
    \left(q_2 + \dfrac{f'(0)}{2k} \right) \gamma_2.
\end{align}

Now the usual boundary value problems of Bleustein-Gulyaev 
wave propagation can be solved. 
For the \textit{metalized boundary condition}, 
$\varphi(0) =0$ and $t_{23}(0) =0$. 
These conditions lead to a homogeneous system of two 
equations for the set of constants $\{\gamma_1, \gamma_2 \}$.  
That set is non-trivial when the following \textit{dispersion 
equation for the metalized boundary condition} is satisfied 
for $v = v_m$ (say),
\begin{equation} \label{metal(i)}
 \left(\dfrac{v_m}{v_T^\circ}\right)^2 = 1 + \dfrac{c_0}{k^2} 
    - \left[ \chi^2 \left(\sqrt{1 + \dfrac{c_0}{k^2}} 
                             + \dfrac{f'(0)}{2k}\right) 
	             - \dfrac{f'(0)}{2k} \right]^2,
\quad 
	\chi^2 := 
  \dfrac{e_{15}^{\circ 2}}
     {c^\circ_{44}\epsilon_{11}^\circ + e_{15}^{\circ 2}}.
\end{equation}
Here, the positive quantity $\chi^2$ is the (bulk) transverse-wave 
electromechanical coupling coefficient.
Recall that the classic Bleustein-Gulyaev wave is non-dispersive for a 
\textit{homogeneous} metalized half-space. 
Its speed $v^\circ_m$ is given by 
\begin{equation} \label{metalHomogeneous}
	(v^\circ_m/v_T^\circ)^2 = 1 - \chi^4.
\end{equation}
Hence the effect of an inhomogeneity of the form 
found in Table 1 is readily seen from the comparison of the 
last two equations. 

For the \textit{free (un-metalized) boundary condition}, 
$t_{23}(0) =0$ and $d_2(0) = \imi k \epsilon_0 \varphi(0)$, 
where $\epsilon_0$ is the permittivity of vacuum
(see for instance (Royer and Dieulesaint, 2000, p. 310).) 
These conditions lead again to a homogeneous system of two 
equations for the set of constants $\{\gamma_1, \gamma_2 \}$.  
The \textit{dispersion equation for the free boundary condition}, 
linking the wave speed $v_f$ (say) to the wave number is now:
\begin{equation} \label{free(i)}
\left(\dfrac{v_f}{v_T^\circ}\right)^2 = 1 + \dfrac{c_0}{k^2} 
  - \left[\chi^2 
           \dfrac{\sqrt{1 + \dfrac{c_0}{k^2}} + \dfrac{f'(0)}{2k}}
                 {1+\dfrac{\epsilon^\circ_{11}}{\epsilon_0} 
                    \left(\sqrt{1 + \dfrac{c_0}{k^2}}
                           + \dfrac{f'(0)}{2k} \right)}  
	                              - \dfrac{f'(0)}{2k} \right]^2.
\end{equation}
Comparison of $v_f$ and $v_m$ shows that $v_f > v_m$, 
whatever the choice of $f$ in Table 1; 
so, by \eqref{q1q2}, the wave penetrates deeper into the 
substrate when its surface is not metalized, as is the case for 
a homogeneous substrate. 
Recall that for a homogeneous half-space, 
the classic Bleustein-Gulyaev wave is non-dispersive for ``free'' 
boundary conditions, and that it travels at speed $v^\circ_f$ 
given by 
\begin{equation} \label{freeHomogeneous}
(v^\circ_f/v_T^\circ)^2 = 
  1 - \chi^4/(1+\epsilon^\circ_{11}/\epsilon_0)^2.
\end{equation}
Note that both $v_f$ and $v_m$ are such that \eqref{condition} is
verified.

In Case \textit{(ii)} (profiles P0 and P6-P10), 
it is found that the attenuation factors are still given by 
\eqref{q1q2}, and that at the interface
\begin{align}
& U_3(0) =  \left( q_1 - \dfrac{f'(0)}{2k} \right) \gamma_1, 
\notag \\
& \varphi(0) =  \dfrac{e_{15}^\circ}{\epsilon_{11}^\circ}
                   \left(q_1 - \dfrac{f'(0)}{2k} \right)\gamma_1 
                   + \left(q_2 - \dfrac{f'(0)}{2k} \right)\gamma_2,
\notag \\
& t_{23}(0) = \imi k
    \left[\left(c^\circ_{44} 
          + \dfrac{e_{15}^{\circ 2}}{\epsilon_{11}^\circ}\right)
     \left(1 - \dfrac{v^2}{v_T^{\circ 2}} \right)\gamma_1 
            + e_{15}^\circ \gamma_2
      \right] ,
\notag \\
& d_2(0) = -\imi k \epsilon_{11}^\circ \gamma_2.
\end{align}
For the \textit{metalized boundary condition}, 
the dispersion equation  is now:
\begin{equation}\label{metal(ii)}
 \chi^2
   \left( 
    \sqrt{1 + \dfrac{c_0}{k^2} - \dfrac{v_m^2}{v_T^{\circ 2}}}
              - \dfrac{f'(0)}{2k}
    \right)
   =  \left(1 - \dfrac{v_m^2}{v_T^{\circ 2}}\right)  
       \left(\sqrt{1 + \dfrac{c_0}{k^2}} - \dfrac{f'(0)}{2k}\right),
\end{equation}
and for the \textit{free (un-metalized) boundary condition}, 
the dispersion equation is now:
\begin{equation} \label{free(ii)}
 \chi^2
   \left( 
    \sqrt{1 + \dfrac{c_0}{k^2} - \dfrac{v_f^2}{v_T^{\circ 2}}}
              - \dfrac{f'(0)}{2k}
    \right)
   =  \left(1 - \dfrac{v_f^2}{v_T^{\circ 2}}\right)  
       \left(\sqrt{1 + \dfrac{c_0}{k^2}} - \dfrac{f'(0)}{2k}
                 +  \dfrac{\epsilon^\circ_{11}}{\epsilon_0}\right).
\end{equation}
These equations could be rationalized but this process 
might introduce spurious speeds. 
It can be checked that they coincide respectively with \eqref{metal(i)}
and \eqref{free(i)} for the exponential inhomogeneity function P0, 
and with \eqref{metalHomogeneous} and \eqref{freeHomogeneous} 
for the homogeneous substrate.

\subsection{Examples}

Consider that the substrate is made of a functionally graded material 
for which the material properties at the interface $x_2=0$ are those 
of a PZT-4 ceramic (Jaffe and Berlincourt, 1965): 
$c^\circ_{44} = 2.56 \times 10^{10}$ N/m$^2$, 
$e^\circ_{15} = 12.7$ C/m$^2$, 
$\epsilon^\circ_{11} = 650 \times 10^{-11}$ F/m, 
$\rho^\circ = 7500$ kg/m$^3$.
The permittivity of vacuum is taken as:
$\epsilon_0 = 8.854 \times 10^{-12}$ F/m. 

When the substrate is \textit{homogeneous}, the Bleustein-Gulyaev wave 
travels with speeds: $v^\circ_m = 2256.85$ m/s and 
$v^\circ_f = 2592.65$ m/s, for metalized and free boundary conditions, 
respectively.
The surface \textit{electromechanical coupling coefficient} $K_S^2$ 
is given by (Royer and Dieulesaint, 2000, p.296),
\be
 K_S^2 = 
  \dfrac{v^{\circ 2}_f - v^{\circ 2}_m}
         {v^{\circ 2}_f 
            + \frac{\epsilon_0}{\epsilon_{11}^\circ}v^{\circ 2}_m}   
       \approx
  \dfrac{v^{\circ 2}_f - v^{\circ 2}_m}
         {v^{\circ 2}_f},
\en 
the latter approximation being justified in the PZT-4 case.
Here, $K_S^2 \approx 0.242$. 

In the first example, the inhomogeneity function is 
\textit{decreasing exponential}: 
$f(x_2) = \exp (-2\beta x_2)$, $\beta >0$ (profile P0 of Table 1). 
Then the mechanical displacement $U_3(x_2)$ varies as: 
$(1/\sqrt{f(x_2)})\exp{-kq_1 x_2} = \exp{-k(q_1 - \beta/k) x_2}$,
and it is found here that 
\be \label{q1Ex1}
q_1 - \dfrac{\beta}{k} = 
 \chi^2\left(\sqrt{1+\dfrac{\beta^2}{k^2}} - \dfrac{\beta}{k} \right),
\quad
\dfrac{\chi^2}
 {\sqrt{1+\dfrac{\beta^2}{k^2}} + \dfrac{\beta}{k} 
     + \dfrac{\epsilon_{11}^\circ}{\epsilon_0}},
\en
for metalized and free boundary conditions, respectively.
Both quantities are clearly positive and the decay is secured. 
The dispersion equations \eqref{metal(i)} and 
\eqref{free(i)} give the wave speed in terms of the dimensionless 
quantity $\beta / k$. 
The range for this quantity is chosen so that the inhomogeneity 
function decreases with depth in a slower fashion 
than the mechanical displacement for the 
metalized boundary condition --- for the free boundary condition, 
the wave speed is so close to the body wave speed 
($v_f^\circ = 0.9999998 v_T$) that the displacement hardly decays at 
all.
In other words, $\beta/k$ satisfies: 
$2\beta / k < q_1 - \beta /k$,
where the right hand-side is given by \eqref{q1Ex1}$_1$. 
This is equivalent to: $\beta/k < \chi^2/(2\sqrt{1+\chi^2}) = 0.2015$. 
Figure 1a shows the variations of $v_m$ (lower curve) and $v_f$ 
(upper curve) with $\beta/k$ from 0 (homogeneous PZT-4 substrate) to 
0.2.
In this range, the inhomogeneity function has no noticeable influence 
on the speed of the Bleustein-Gulyaev wave 
with free boundary conditions, 
whereas it slows down significantly the Bleustein-Gulyaev wave 
with metalized boundary conditions, resulting in an increasing
electromechanical coupling coefficient $K_S^2$ (Figure 1b), from 
0.242 to 0.324.  

For the second example, the inhomogeneity function is 
\textit{inverse quadratic}: 
$f(x_2) = 1/(\beta x_2 + 1)^2$, $\beta >0$ (profile P6 of Table 1). 
Then the mechanical displacement $U_3(x_2)$ varies as: 
$(1/\sqrt{f(x_2)})\exp{-kq_1 x_2} = (\beta x_2 + 1) \exp{-k q_1 x_2}$.
Here the decay is secured when \eqref{condition} is satisfied,
which is equivalent to: $v < v_T$, the same condition as in the 
homogeneous substrate.
For this profile, $c_0=0$ and $f'(0) = -2\beta$, so that the  
dispersion equations \eqref{metal(ii)} and \eqref{free(ii)} are 
easily solved.
The metalized boundary condition gives:
\begin{equation} 
 \left(\dfrac{v_m}{v_T^\circ}\right)^2 = 
  1 - \dfrac{\chi^4}{4\left(1 + \dfrac{\beta}{k}\right)^2} 
         \left[1+ \sqrt{1 + \dfrac{4}{\chi^2} 
                 \left(1+ \dfrac{\beta}{k}\right)\dfrac{\beta}{k}} 
	              \right]^2,
\end{equation}
and the free boundary condition gives:
\begin{equation} 
 \left(\dfrac{v_f}{v_T^\circ}\right)^2 = 
  1 - \dfrac{\chi^4}{4\left(1 + \dfrac{\beta}{k} 
                  + \dfrac{\epsilon_{11}^\circ}{\epsilon_0}\right)^2} 
         \left[1+ \sqrt{1 + \dfrac{4}{\chi^2} 
                 \left(1+ \dfrac{\beta}{k}
                         + \dfrac{\epsilon_{11}^\circ}{\epsilon_0}
                            \right)\dfrac{\beta}{k}} 
	              \right]^2,
\end{equation}
For the purpose of comparison with the first example, 
Figures 2a and 2b display the variations of the wave speeds with 
the dimensionless quantity $\beta / k$ over the same range 
$0 \le \beta/k \le 0.2$. 
They show that the influence of each inhomogeneity functions 
is very much the same: here, the electromechanical coupling 
coefficient increases (from 0.242) to 0.310 instead of 0.324 
for the decreasing exponential profile. 
The present inverse quadratic inhomogeneity is however more 
satisfying to consider from a ``physical'' point of view, 
because it decreases slower with depth than an exponential 
inhomogeneity, and it can thus describe a situation where the 
wave is confined near the surface while the material parameters 
\eqref{inhomogeneityLAw} vanish at a greater distance. 

The next Section presents a third example of inhomogeneity function, 
this time yielding a profile for which the material parameters neither
vanish nor blow-up with distance from the interface. 

\section{An asymptotically homogeneous half-space}

Consider the well-known solution 
$[\vec{U}(x_2), \vec{V}(x_2)]^T$ (say) to the piezoacoustic equations 
\eqref{1stOrder} in a \textit{homogeneous} substrate:
\begin{equation} \label{homSoln}
 \begin{bmatrix} 
  \vec{U}(x_2) \\ \vec{V}(x_2)
     \end{bmatrix} = 
  \gamma_1  \text{e}^{-k  \eta x_2} 
     \begin{bmatrix} 
        1 \\ 
        e_{15}^\circ / \epsilon_{11}^\circ\\
        \imi k\eta
         (c^\circ_{44} 
           + e_{15}^{\circ 2} / \epsilon_{11}^\circ)\\
          0
     \end{bmatrix}
  + \gamma_2  \text{e}^{-k x_2} 
        \begin{bmatrix} 
          0 \\ 1 \\
          \imi k e_{15}^\circ \\
          -\imi k \epsilon_{11}^\circ
        \end{bmatrix},
\quad 
 \eta:= \sqrt{1 - \left(\frac{v}{v_T^\circ}\right)^2}.
\end{equation}
These functions satisfy \eqref{1stOrder} when $f \equiv 1$ that is,
\begin{equation} \label{homogeneousSolution}
  \vec{U}' = \imi \vec{N_2 V}, \quad \vec{V}' = \imi k^2 \vec{K U}.
\end{equation}
 
Now seek a solution $[\vec{u}, \vec{v}]^T$ to the piezoacoustic 
equations \eqref{1stOrder} for the functionally graded material 
in the form:
\begin{equation} \label{N=1}
  \vec{u} = \dfrac{u_0}{k} \vec{U}' + u_1 \vec{U}, 
  \quad 
  \vec{v} = \dfrac{v_0}{k} \vec{V}' + v_1 \vec{V},
\end{equation}
where $u_0$, $u_1$, $v_0$, $v_1$ are yet unknown scalar functions 
of $x_2$. 
By differentiation and substitution, it is found that if they satisfy 
\begin{equation} 
  u_0 = v_0/f, \quad
  u_0'/k + u_1 = v_1/f, \quad
  v_0'/k + v_1 = fu_1, \quad
  u_1' = 0, \quad
  v_1' = 0,
\end{equation}
then \eqref{1stOrder} is satisfied. 
The choice 
\begin{equation} \label{choice}
	u_0 = 1/\sqrt{f}, \quad 
	v_0 = \sqrt{f}, \quad
	u_1 = -\dfrac{\beta}{k} \tanh \delta, \quad
	v_1 = -\dfrac{\beta}{k \tanh \delta},
\end{equation}
(where $\beta$, $\delta$ are constants) takes care of 
\eqref{1stOrder}$_{1,4,5}$. 
Then \eqref{1stOrder}$_{2,3}$ both reduce to 
\begin{equation}
	(\sqrt{f})' - (\beta/\tanh \delta) = -(\beta \tanh \delta) f,
\end{equation}
a solution of which is 
\begin{equation} \label{tanh^2}
	f(x_2) = \dfrac{\tanh^2 (\beta x_2 + \delta)}{\tanh^2 \delta}.
\end{equation}
This inhomogeneity function has the sought-after property of never 
vanishing (if $\delta>0$) nor blowing-up as $x_2$ spans the whole 
half-space occupied by the substrate. 
It describes a material for which the parameters 
$c_{44}(x_2)$, $e_{15}(x_2)$, $\epsilon_{11}(x_2)$, $\rho(x_2)$  
chan\-ge smoothly from their initial values 
$c^\circ_{44}$, $e^\circ_{15}$, $\epsilon^\circ_{11}$, 
$\rho^\circ$ 
at the interface $x_2=0$ to a higher asymptotic value 
$\{c^\circ_{44}$, $e^\circ_{15}$, $\epsilon^\circ_{11}$, $\rho^\circ \}
 /\tanh^2 \delta$, where $\delta >0$ is an adjustable parameter.
The parameter $\beta$ can also be adjusted to describe a not only a 
slow but also a rapid variation with depth, which would confine the 
inhomogeneity to a thin layer near the surface.
This latter opportunity was excluded with the profiles of Section 3, 
because of their blow-up or vanishing behaviors.

Now the substitution of $f$ into \eqref{choice} and then into the 
solution \eqref{N=1}, leads to the following expressions for the 
fields at the interface,
\begin{align}
& U_3(0) =  \left(\eta + \frac{\beta}{k}\tanh\delta \right)\gamma_1, 
\notag \\
& \varphi(0) =  
   \frac{e_{15}^\circ}{\epsilon_{11}^\circ}
        \left(\eta + \frac{\beta}{k}\tanh\delta \right)\gamma_1 
               + \left(1 + \frac{\beta}{k}\tanh\delta \right)\gamma_2,
\notag \\
& t_{23}(0) = \imi k
    \left[\left( c^\circ_{44} 
          + \dfrac{e_{15}^{\circ 2}}{\epsilon_{11}^\circ}\right)
     \eta \left(\eta + \frac{\beta}{k\tanh\delta} \right)\gamma_1 
    + e_{15}^\circ\left(1 + \frac{\beta}{k\tanh\delta}\right)\gamma_2
      \right] ,
\notag \\
& d_2(0) = -\imi k \epsilon_{11}^\circ
    \left(1 + \frac{\beta}{k\tanh\delta} \right) \gamma_2.
\end{align}

The \textit{dispersion equation for the metalized boundary condition} 
is a quadratic in $\eta = \sqrt{1 - (v_m/v_T^{\circ})^2}$:
\begin{equation}
  \eta \left(\eta + \frac{\beta}{k\tanh\delta} \right)
           \left(1 + \frac{\beta}{k}\tanh\delta \right)      
 - \chi^2  \left(\eta + \frac{\beta}{k}\tanh\delta \right)
                   \left(1 + \frac{\beta}{k\tanh\delta}\right)
    = 0.       
\end{equation}
At $\beta = 0$, the function $f$ in \eqref{tanh^2} is that of a 
homogeneous substrate ($f \equiv 1$), 
and this equation gives: $\eta = \chi^2$, 
which, once squared, is \eqref{metalHomogeneous}.

The \textit{dispersion equation for the free boundary 
condition} is also a quadratic, now in 
$\eta = \sqrt{1 - (v_f/v_T^\circ)^2}$:
\begin{equation}
  \eta \left(\eta + \frac{\beta}{k\tanh\delta} \right)
       \left[ \dfrac{1 + \dfrac{\beta}{k}\tanh\delta}
                     {1 + \dfrac{\beta}{k\tanh\delta}} 
                + \dfrac{\epsilon^\circ_{11}}{\epsilon_0}
       \right]      
 - \chi^2  \left(\eta + \frac{\beta}{k}\tanh\delta \right)
    = 0.       
\end{equation}
At $\beta = 0$,  this equation gives: 
$\eta(1 + \epsilon^\circ_{11}/\epsilon_0) = \chi^2$, 
which, once squared, is \eqref{freeHomogeneous}.

Each dispersion equation is a quadratic giving a priori two roots: 
one  tends to $\eta = 0$ (and so $v = v_T$) as $\beta \rightarrow 0$ 
(homogeneous substrate) and can be ruled out.

For the third example, consider an inhomogeneous substrate whose 
properties increase continuously according to  \eqref{tanh^2}
from those of a PZT-4 ceramic at the interface $x_2=0$ to asymptotic 
values which are 10\% greater 
($\tanh \delta = 1/\sqrt{1.1} \approx 0.953$).
Figure 3 shows the variations of the inhomogeneity function 
with depth, for several values of the parameter $\beta/k$: 
clearly for $\beta /k >2$, the inhomogeneity is confined  within a 
layer near the surface whose thickness is less than a wavelength. 
Figure 4a shows the variations of $v_m$ (lower curve) and $v_f$ 
(upper curve) with $\beta/k$ from 0 (homogeneous PZT-4 substrate) to 2.
In this range, the inhomogeneity function again 
has no noticeable influence on the speed of the Bleustein-Gulyaev wave 
with free boundary conditions.  
However  here the speed of the Bleustein-Gulyaev wave 
with metalized boundary conditions is \textit{always greater} than 
in the homogneous substrate, 
resulting in a smaller electromechanical coupling coefficient 
$K_S^2$ (Figure 4b).

\section{More inhomogeneity functions}

The previous Section presented a method to derive an inhomogeneity 
function for which exact Bleustein-Gulyaev solutions are possible, 
but which did not rely on finding governing equations with 
constant coefficients as in Section 3.
That method is due in essence to Varley and Seymour (1988) 
(see also (Erdogan and Ozturk, 1992)) and it can be generalized 
to yield an infinity of such inhomogeneity functions. 

First, seek a solution 
$[\vec{u}, \vec{v}]^T$ to the piezoacoustic equations 
\eqref{1stOrder} in the form:
\begin{equation}
  \vec{u} = 
 \sum_{n=0}^p \dfrac{u_n}{k^{p-n}} \vec{U}^{(p-n)}, 
  \quad 
  \vec{v} = \sum_{n=0}^p \dfrac{v_n}{k^{p-n}} \vec{V}^{(p-n)},
\end{equation}
(thus \eqref{N=1} corresponds to $p=1$), where $u_n$, $v_n$ 
($n = 0, \ldots, p$) are unknown functions of $x_2$. 
Differentiate once, substitute into \eqref{1stOrder}, and use 
\eqref{homogeneousSolution} to get
\begin{equation}
   \dfrac{1}{k^p}\left(u_0 - \dfrac{1}{f}v_0\right) \vec{U}^{(p-1)} 
 +  \sum_{n=1}^p \dfrac{1}{k^{p-n}}
     \left(\dfrac{u'_{n-1}}{k} + u_n - \dfrac{1}{f}v_n \right)
                                               \vec{U}^{(p-n-1)}
 + u'_p \vec{U} = \vec{0},
\end{equation}
and 
\begin{equation}
   \dfrac{1}{k^p}\left(v_0 - f u_0\right) \vec{V}^{(p-1)} 
 +  \sum_{n=1}^p \dfrac{1}{k^{p-n}}
     \left(\dfrac{v'_{n-1}}{k} + v_n - f u_n \right)
                                               \vec{V}^{(p-n-1)}
 + v'_p \vec{V} = \vec{0}.
\end{equation}
Both differential equations are satisfied when the following set of  
equations is satisfied,
\begin{align} 
& u_0 = 1/\sqrt{f}, \quad v_0 = \sqrt{f}, \quad
  u_p' = 0, \quad
  v_p' = 0,
\notag \\
&  u_{n-1}'/k + u_n = v_n/f, \quad
  v_{n-1}'/k + v_n = f u_n.
\end{align}

Varley and Seymour (1988) found an infinity of $f$ such that this set 
can be completely solved.
Because there is little value in reproducing their derivations,
the reader is referred to their article for explicit examples.
It suffices to notice that in general the solutions are combinations 
of trigonometric and/or hyperbolic functions, and that a great number 
of arbitrary constants are at disposal for curve fitting. 
For instance at $p=2$, Varley and Seymour present at least 16 possible 
forms for $f$, each involving 5 arbitrary constants.
As another example, it can be checked directly here 
that the polynomial function 
\begin{equation}
  f(x_2) = (\beta x_2 + 1)^{2p},
\end{equation}
where $p$ is an integer, is also suitable; then, 
\begin{equation}
  u_n = b_n (\beta x_2 + 1)^{-(p+n)}, \quad 
  v_n = a_n (\beta x_2 + 1)^{p-n}, \quad (n = 0, \ldots, p),
\end{equation}
where $a_n$ and $b_n$ are determined by recurrence from 
$a_0 = b_0 = 1$
and 
\begin{equation}
 a_n = - \dfrac{\beta}{2kn}(p-n+1)(p+n)a_{n-1}, 
    \quad  b_n = \dfrac{p-n}{p+n}a_n,  \quad (n = 1, \ldots, p).
\end{equation}




\newpage
\thispagestyle{plain}

\begin{center}
Table 1.
\textit{Some simple inhomogeneity functions}

\rule[-3mm]{0mm}{8mm} 

\noindent
\begin{tabular}{l | c | c | c |}
 & $c_0$  & $p(x_2)$ & $f(x_2)$ 
\\
\hline

\rule[-1mm]{0mm}{6mm} 

P0  & $\beta^2$ & $\pm 2 \beta$ & $\exp(\pm 2 \beta x_2)$

\\

\hline 
\hline 

\rule[-1mm]{0mm}{6mm} 

P1  &  0  & $2 \beta /(\beta x_2 + 1)$ & $(\beta x_2 + 1)^2$

\\

\hline 

\rule[-4mm]{0mm}{12mm} 

P2  & $\beta^2$ & $2 \beta \tanh(\beta x_2 + \delta) $ 
    & $\dfrac{\cosh^2(\beta x_2 + \delta)}{\cosh^2 \delta}$

\\

\hline 

\rule[-4mm]{0mm}{12mm} 

P3  & $\beta^2$ & $2 \beta / \tanh(\beta x_2 + \delta)$ 
    & $\dfrac{\sinh^2(\beta x_2 + \delta)}{\sinh^2 \delta}$

\\

\hline 

\rule[-4mm]{0mm}{12mm} 

P4  & $-\beta^2$ & $-2 \beta \tan(\beta x_2 + \delta) $ 
    & $\dfrac{\cos^2(\beta x_2 + \delta)}{\cos^2 \delta}$

\\

\hline 

\rule[-4mm]{0mm}{12mm} 

P5  & $-\beta^2$ & $2 \beta / \tan(\beta x_2 + \delta) $ 
    & $\dfrac{\sin^2(\beta x_2 + \delta)}{\sin^2 \delta}$

\\

\hline 
\hline 

\rule[-4mm]{0mm}{12mm} 

P6  &     0     & $-2 \beta /(\beta x_2 + 1)$ 
    & $\dfrac{1}{(\beta x_2 + 1)^2}$

\\

\hline 

\rule[-4mm]{0mm}{12mm} 

P7  & $\beta^2$ & $-2 \beta \tanh(\beta x_2 + \delta) $ 
    & $\dfrac{\cosh^2 \delta}{\cosh^2(\beta x_2 + \delta)}$

\\

\hline 

\rule[-4mm]{0mm}{12mm} 

P8  & $\beta^2$ & $-2 \beta / \tanh(\beta x_2 + \delta) $ 
    & $\dfrac{\sinh^2 \delta}{\sinh^2(\beta x_2 + \delta)}$

\\

\hline 

\rule[-4mm]{0mm}{12mm} 

P9  & $-\beta^2$ & $2 \beta \tan(\beta x_2 + \delta) $ 
    & $\dfrac{\cos^2 \delta}{\cos^2(\beta x_2 + \delta)}$

\\

\hline 

\rule[-4mm]{0mm}{12mm} 

P10  & $-\beta^2$ & $-2 \beta / \tan(\beta x_2 + \delta) $ 
     & $\dfrac{\sin^2 \delta}{\sin^2(\beta x_2 + \delta)}$

\\

\hline 

\end{tabular}
\end{center}


\newpage

\begin{figure}
 \centering 
  \mbox{\subfigure{\epsfig{figure=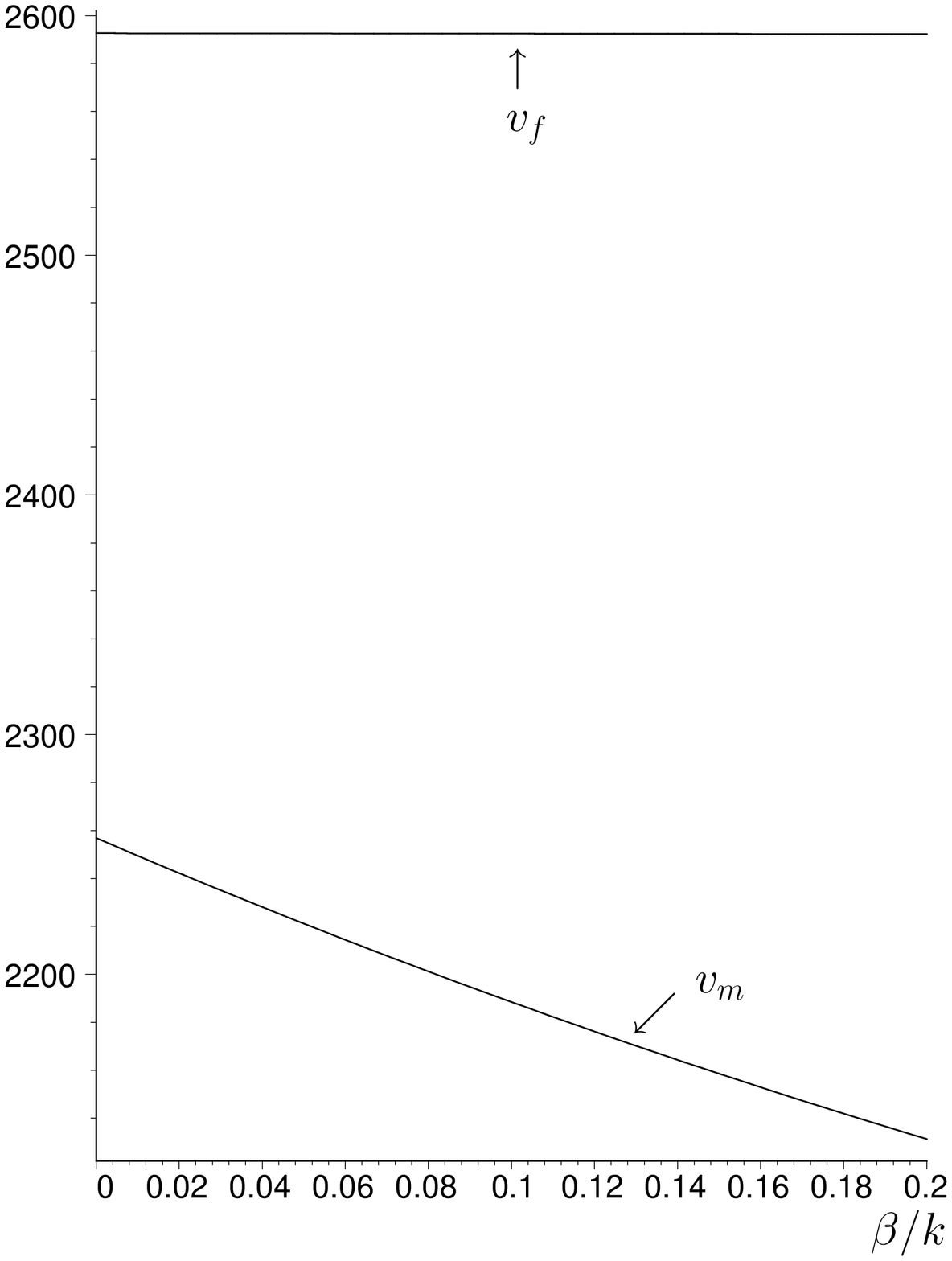,
width=.45\textwidth}}
  \quad \quad
     \subfigure{\epsfig{figure=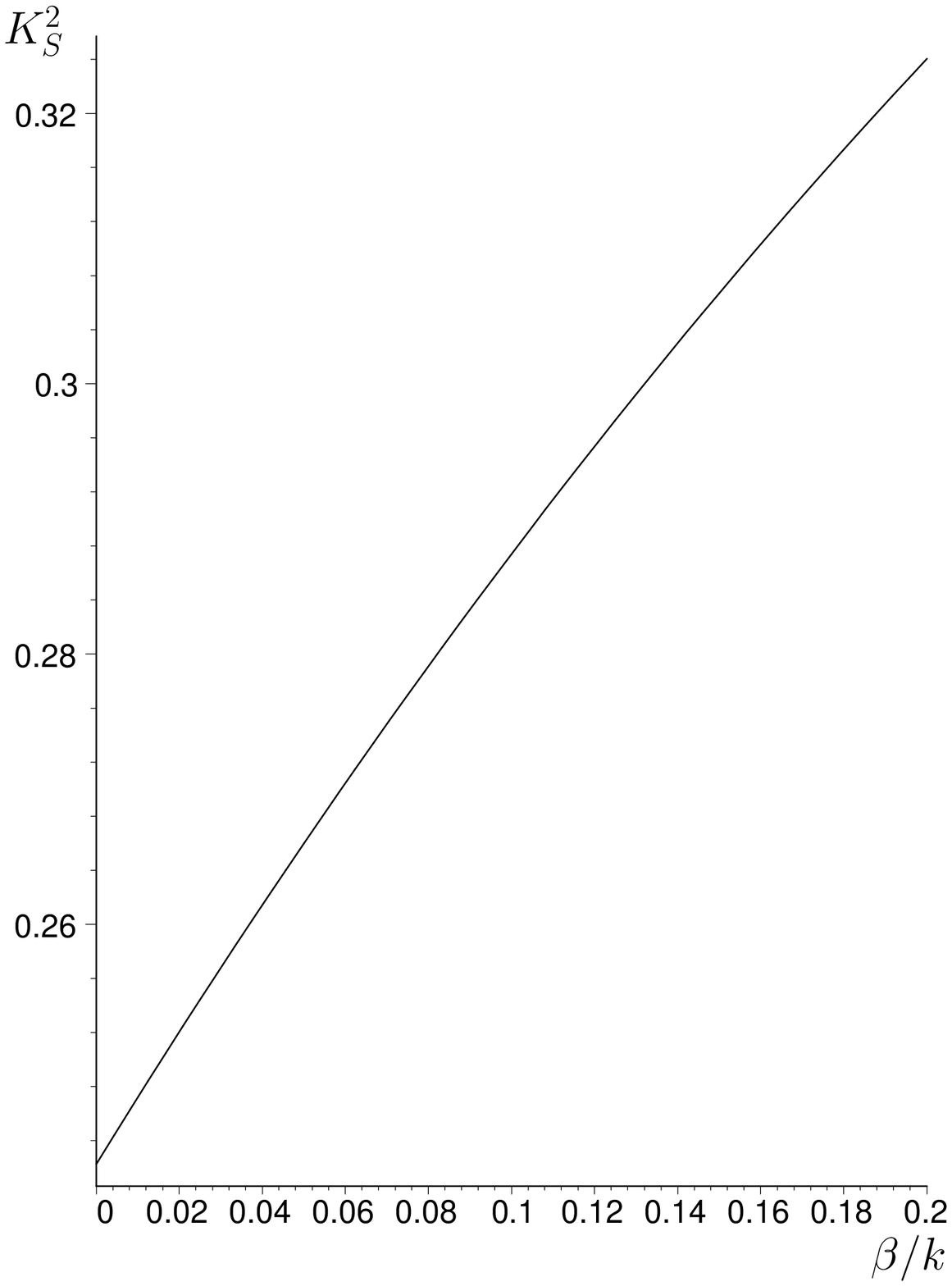,
width=.45\textwidth}}}
\caption{Influence of a decreasing exponential inhomogeneity function 
on the wave speed (free and metalized boundary conditions) and on the 
electromechanical coupling coefficient.}
\end{figure}


\newpage

\begin{figure}
 \centering 
  \mbox{\subfigure{\epsfig{figure=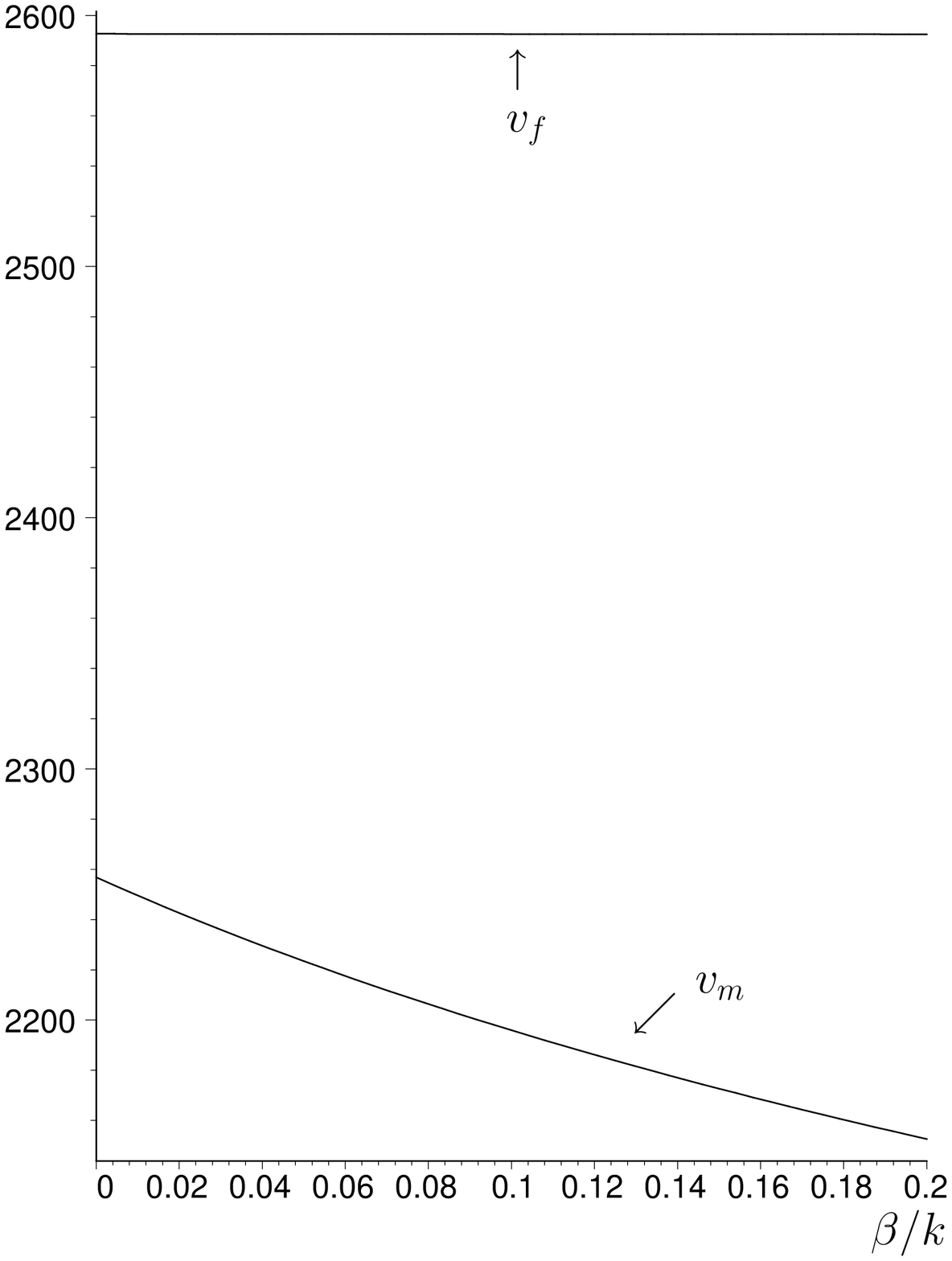,
width=.45\textwidth}}
  \quad \quad
     \subfigure{\epsfig{figure=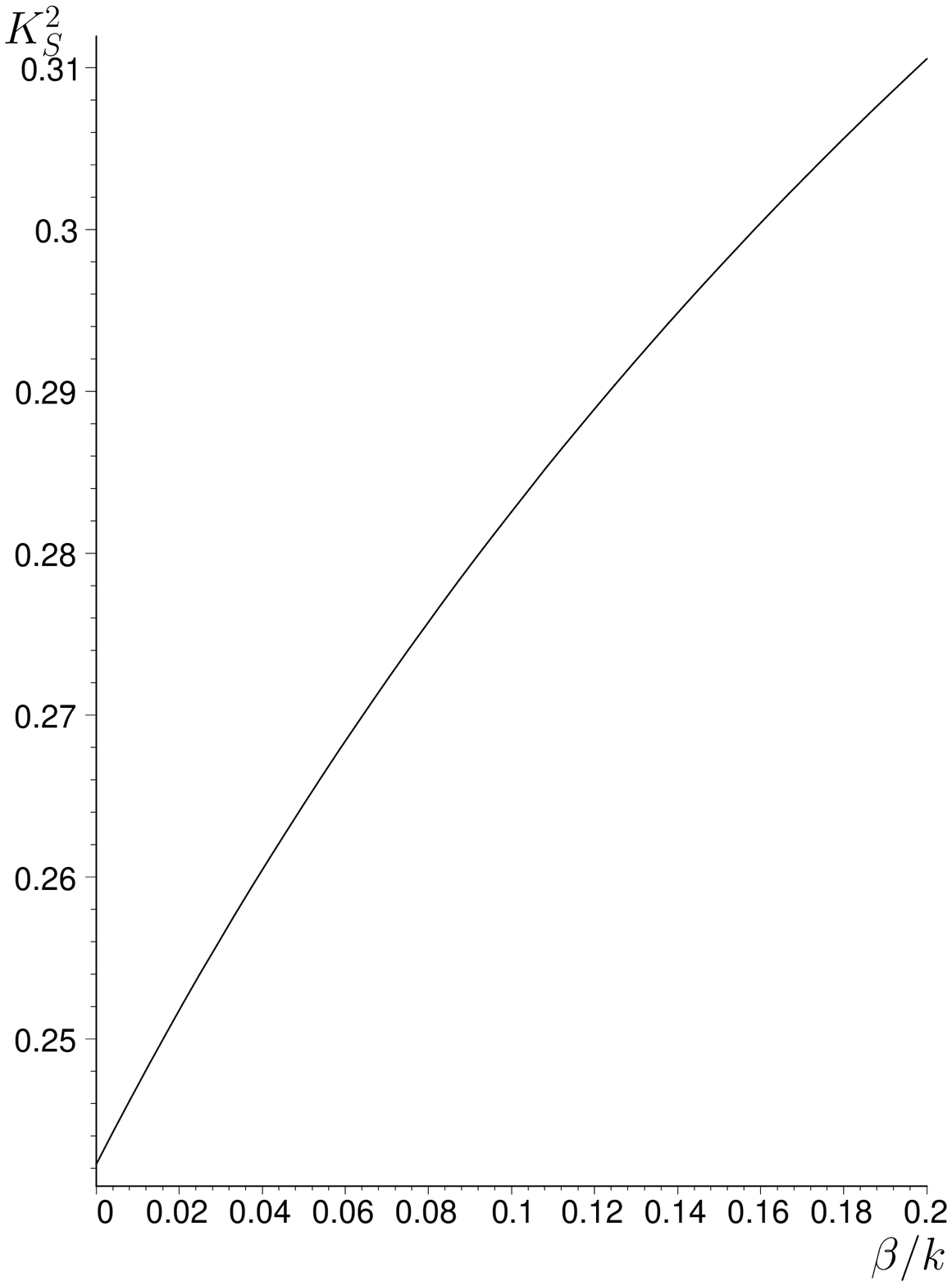,
width=.45\textwidth}}}
\caption{Influence of an inverse quadratic inhomogeneity function 
on the wave speed (free and metalized boundary conditions) and on the 
electromechanical coupling coefficient.}
\end{figure}


\newpage

\begin{figure}
 \centering 
  \epsfig{figure=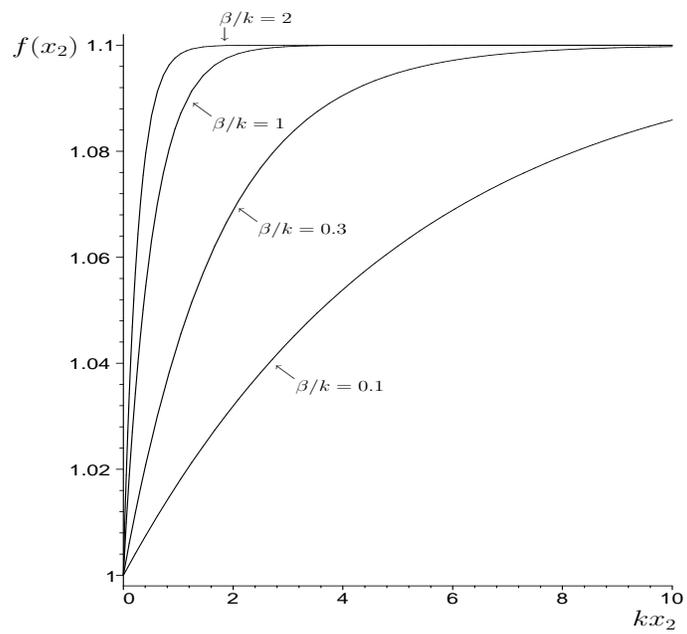, height=.6\textwidth,width=.65\textwidth}
\caption{Variation of an asymptotically homogeneous profile with 
depth for four different values of the dispersion parameter $\beta/k$.}
\end{figure}


\newpage

\begin{figure}
 \centering 
  \mbox{\subfigure{\epsfig{figure=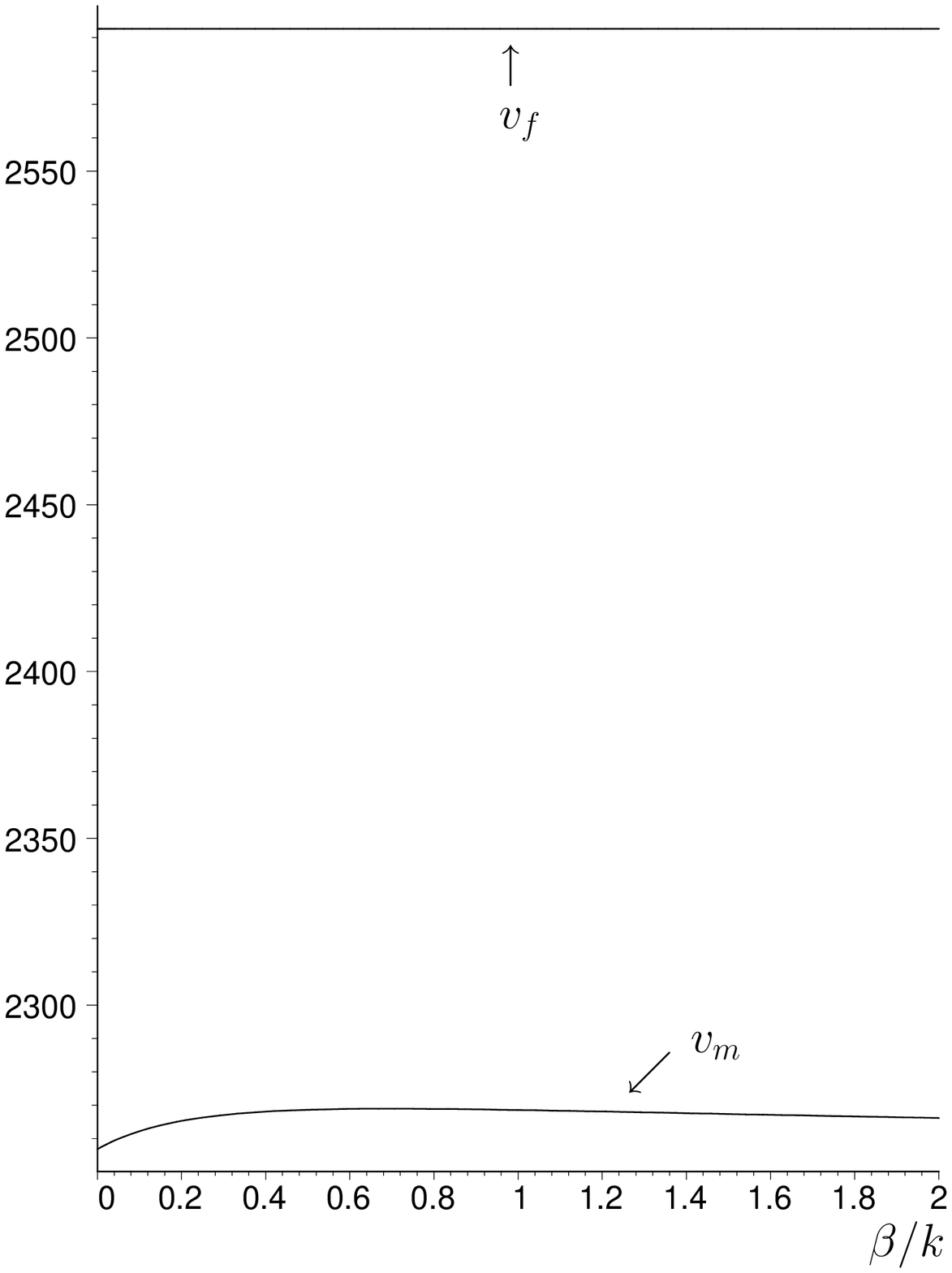,
width=.45\textwidth}}
  \quad \quad
     \subfigure{\epsfig{figure=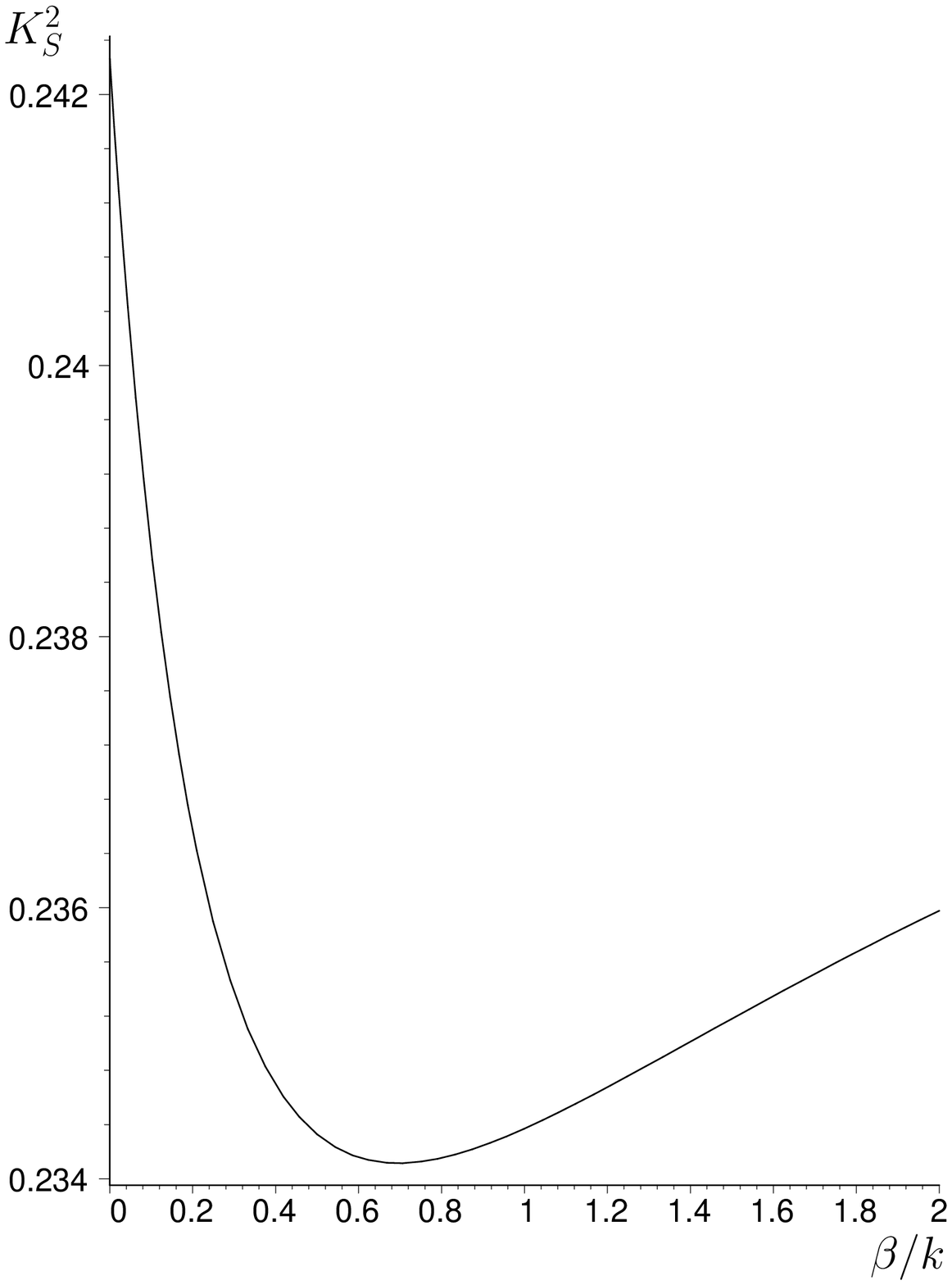,
width=.45\textwidth}}}
\caption{Influence of an asymptotically homogeneous profile 
on the wave speed (free and metalized boundary conditions) and on the 
electromechanical coupling coefficient.}
\end{figure}


\end{document}